# Origins of Standing Stone Astronomy in Britain: new quantitative techniques for the study of archaeoastronomy.

## Gail Higginbottom[a] and Roger Clay[b]

[a.] **Corresponding author.** Department of Physics, School of Physical Sciences, University of Adelaide, Adelaide, South Australia, Australia 5005, and School of Archaeology & Anthropology, The Australian National University, Canberra, ACT. Australia, 0200.  gail.higginbottom@anu.edu.au
[b.] Department of Physics, School of Physical Sciences, University of Adelaide, Adelaide, South Australia, Australia, 5005. roger.clay@adelaide.edu.au

## Abstract

By c. 3000 BC, in the late Neolithic, there had been a significant change in the way people materialized their cosmology across Scotland with the introduction of free-standing stones that continued to be erected almost until the end of the Bronze Age (Burl 2000, 1993). Significantly, a series of astronomical patternings have been empirically verified for many Bronze Age monuments that were erected between 1400-900 BC (Higginbottom *et al.* 2013, 2001, 2000). Further, two series of complex landscape patternings associated with the monuments and their orientations have been identified (Higginbottom *et al.* 2013). However, when and where these patterns were *first* associated with standing-stone structures was unknown. *Through innovative statistics and software* we show that visible astronomical-landscape variables found at Bronze Age sites on the inner isles and mainland of western Scotland were actually *first established* nearly two millennia earlier, with the erection of the first 'great circles' in Britain: Callanish on the Isle of Lewis and Stenness on the Isle of Orkney. In particular, we introduce our new statistical test that enables the *quantitative* determination of astronomical connections of stone circles. It is seen that whilst different standing-stone monuments were created over time (Burl 2000, 1993; Higginbottom *et al.* 2013; Sheridan & Brophy 2012) with a mixture of landscape variables (Higginbottom *et al.* 2013), we nevertheless see that highly relevant aspects *remained unchanged* through these years. This suggests that there is some continuity of this cosmological system through time, despite the various radical material and social changes that occurred from the late Neolithic to the Late Bronze Age (Lynch 2000; Mullin 2001; Owoc 2001).

## 1. Archaeological Background

During the Late Neolithic, the Orkney regional group built the following: the settlements of Skara Brae and Barnhouse, Maeshowe-type chambered cruciform passage tombs, and carved spiral motifs, as well as the large Ness of Brodgar 'temple complex', followed closely by the henge and great circle of the Stones of Stenness (Fig. 1 & Fig 2a). The nearby single standing stones (SStS) are not yet dated. We note that neither such a temple complex nor an associated settlement has yet been found for western Scotland. However, on the west coast of Isle of Lewis, there is a complex of at least eight stone settings, including circles, visually dominated by the great Stone Circle of Callanish with its thin slab-like menhirs, 3-4, m in height and a central monolith, 5 m tall (Callanish 1). Three stone rows (SR) and a stone avenue (SA) extend from the circle (Fig. 2b; Burl 1993: 148-151). Also on Lewis, are at least a further 30 standing-stone sites (StS) of prehistoric origin identifiable today.

Most of the circles are associated with small internal cairns. Whilst we do not yet know when the surrounding circles and StS near Callanish 1 were erected, nor the associated SR and SA, its circle and central monolith were built around 3000-2900 BC and its internal tomb during the early Bronze Age (Ashmore, *in press*; Sheridan & Brophy 2012: 76). These date ranges are inferred through the combination of the order or sequence of erection events at Callanish 1 and 32 radiocarbon-dated samples from excavation (Ashmore, *in press*; Ashmore 1999; Royal Commission of Ancient & Historical Monuments (RCAHMS) – CANMORE digital database). Now Stenness, its slabs measuring 4.57 to 5.2 m high, has a date range for the henge and its hearth of 3100-2650 BC (Burl 2000: 211; Higginbottom *et al.* 2013: 3; Ritchie 1976: 10, Appendices; RCAHMS – CANMORE - under site name). When looking at all of the features of this site, Schulting *et al.*'s Bayesian model of dates shows that it was likely built between 3000-2900 BC (Schulting *et al.* 2010: 35–36). This means that Callanish was built in very close chronological proximity to Stenness. This, and their similar structural elements relating to circularity and to the dead (Higginbottom *in preparation*, Connecting the Great Stone Circles of Scotland (GH1); Higginbottom *et al.* 2013; Richards (a) & (b)) are relevant for us here.

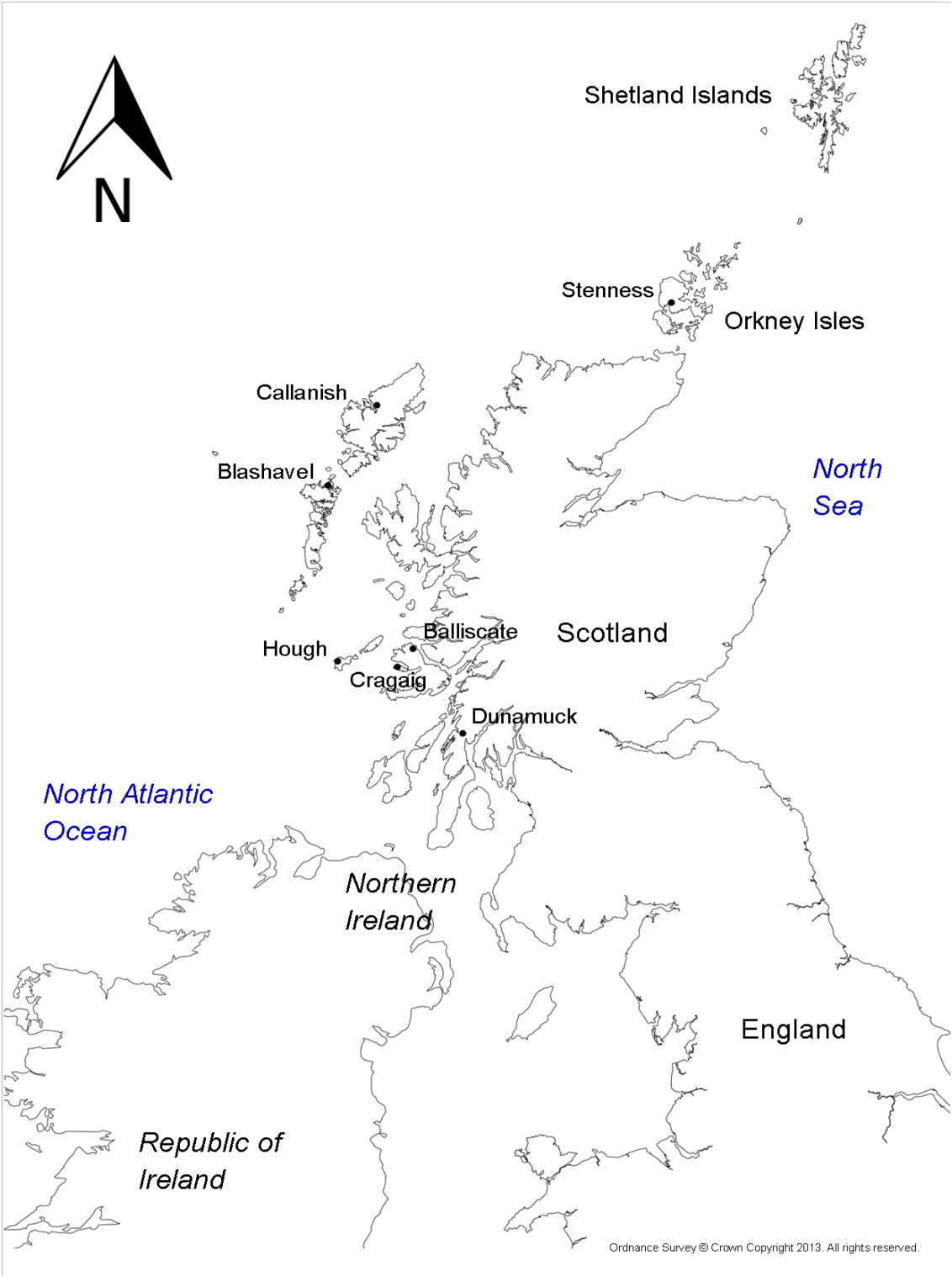

**Fig. 1: Map of region and sites mentioned in text placed on map.** Scale 1:350,000; north arrow indicates grid north. Modified Ordnance Survey map: Ordnance Survey © Crown Copyright 2013. All rights reserved.

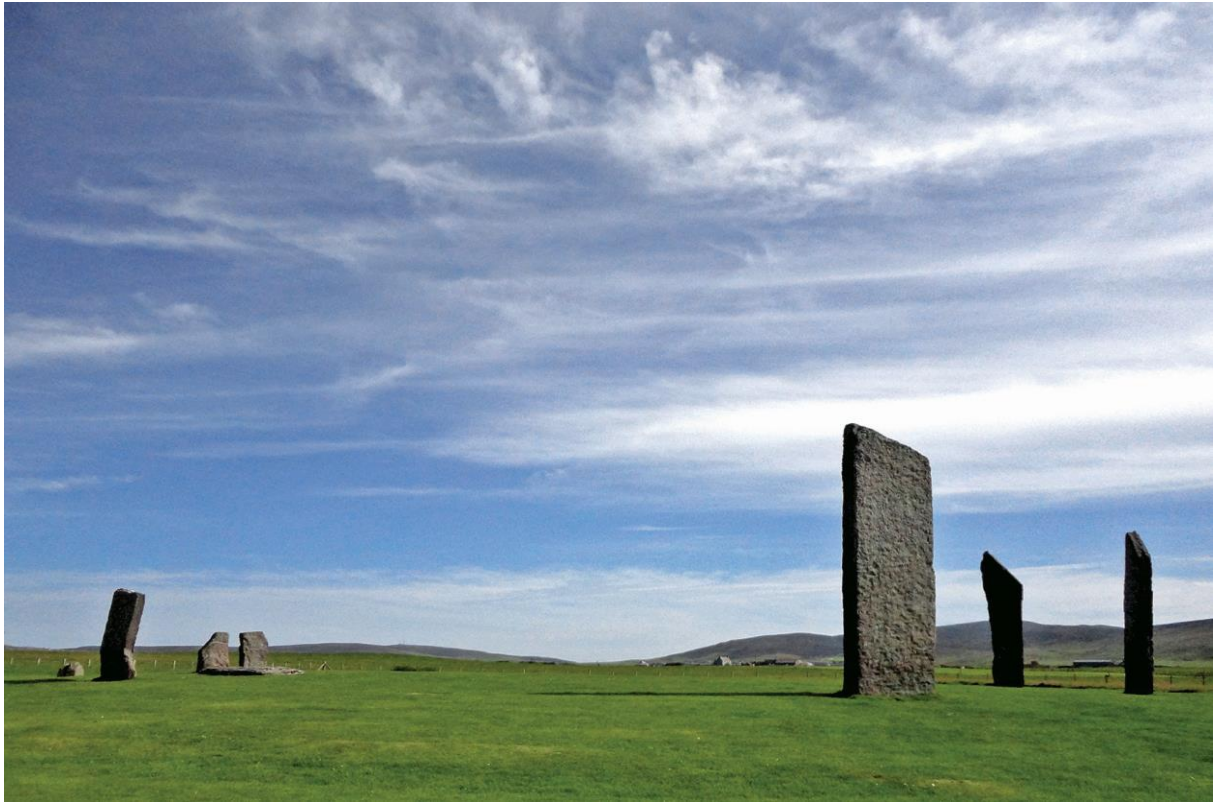

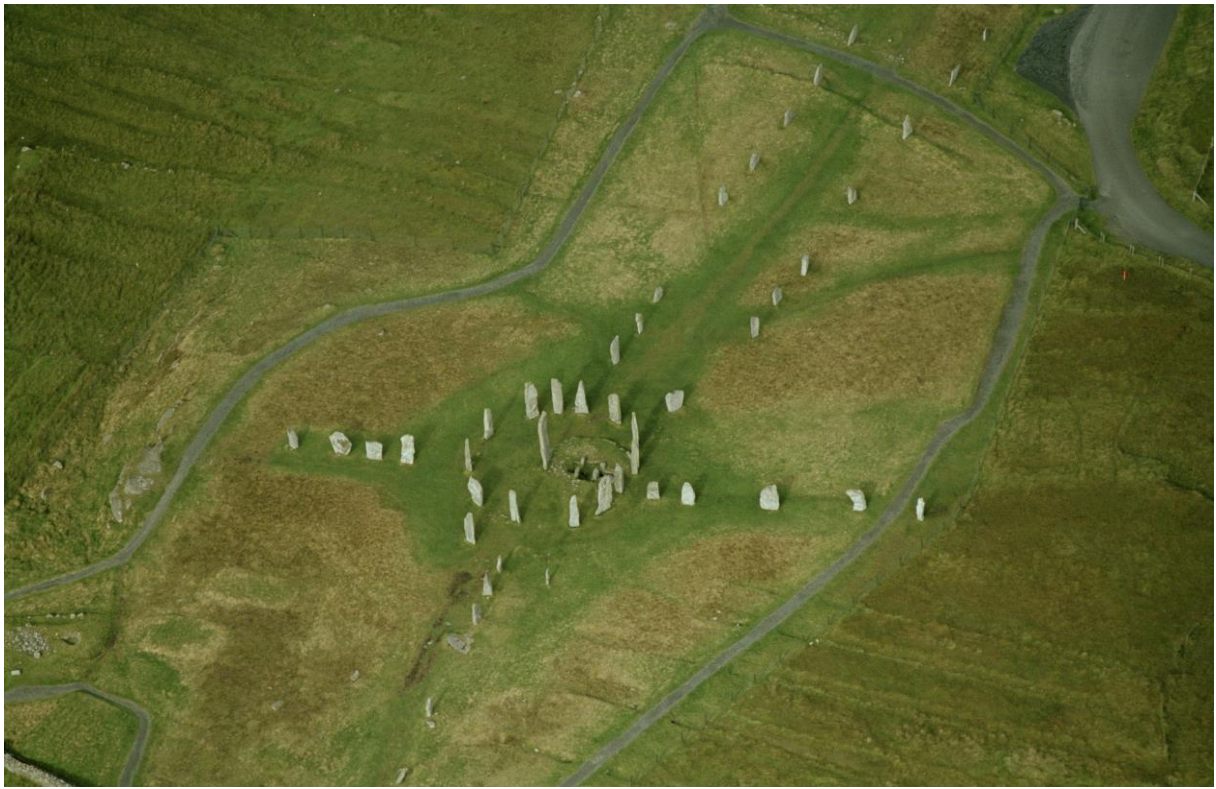

**Fig. 2: The great circles.** A. (top), Stenness, Mainland, Orkney. The circle of the Stones of Stenness is 32.2m by 30.6m (Burl 2010: 210). Its earthen henge is 45 m in diameter, over 7 m wide and over 2 m deep and the circumference is 141.37 m (Burl 1976: 210; Ritchie and Ritchie 1991: 47-50). Photograph Douglas Scott, © Douglas Scott. B. (bottom), Callanish I on Lewis is 13 m in diameter with a long stone avenue running north-southwards (southwards is towards the circle) and single long stone rows radiating outwards towards the other three cardinal points. SC_1023422, © RCAHMS (Aerial Photography Collection). Licensor www.rcahms.gov.uk

We know too, that from the Neolithic to the Bronze Age, the two most common elements associated with standing stones in general across Scotland are that of the dead (usually cremated remains at the base or next to the StS) and astronomical phenomena (Barber 1977-1978: 107; Duffy 2007: 54; Higginbottom *et al.* 2013: 4-15, 17-22; MacKie 2001; Martlew & Ruggles 1996; Richards (a) in Richards 2013; Thom 1971). Through a statistical reassessment of the works of Ruggles and colleagues (e.g. Patrick and Freeman 1985; Ruggles 1984; Ruggles & Martlew, 1992; Ruggles *et al.* 1991) it was shown that a greater number of standing stone monuments than previously thought were deliberately orientated to either the Sun or the Moon across Mull, Coll, Tiree and Argyll, and those findings were extended to Islay and Jura in western Scotland (Higginbottom 2003 *Interdisciplinary Study of Megalithic Monuments in Western Scotland*; PhD thesis, University of Adelaide: 126-132 (Higginbottom 2003; Higginbottom *et al.* 2001; Higginbottom & Clay 1999). Further, sites in areas like Kintyre, Uist and Lewis also appear to contain these same phenomena (GH1 *in preparation*; Higginbottom 2003; Higginbottom *in preparation* Cosmology to the ends of the earth: the Outer Hebrides: (GH2). The range of dates for the linear settings, pairs or short rows in particular, and their possible associated monuments appears to be the mid-late Bronze Age (e.g. Ardnacross, Mull (pair of stone rows): 1250 and 900 cal BC, Martlew & Ruggles 1996: 126 and Ballymeanoch, Argyll – a stone row dated by the single StS next to it: 1370-1040 cal BC; Barber 1977–1978: 107; RCAHMS - Canmore: sample ID GrA-28613). After the Late Neolithic and early Bronze Age, other single standing stones (SS) are dated to a similar late period: Dunure Road in Ayrshire, 1310–1050 cal BC (Duffy 2007: 53; Higginbottom 2013: 7, 15-18). The finds that date the StS are usually cremated human bone from deposits set against the base of the standing stone often sealed under packing stones, like at Ballymeanoch.

We will now provide some necessary information on astronomy before proceeding with the discussions on our quantitative archaeoastronomical research in Scotland, this will be followed by our related landscape archaeological work.

## 2. Understanding basic observational astronomical information

The celestial sphere (CS) is an imaginary sphere as viewed from earth against which the celestial bodies appear to be projected. The CS's axis is the extension of the earth's own axis. The apparent dome of the visible sky forms half of this sphere. The position of an object on the celestial sphere is specified by its equatorial coordinates: right ascension and declination. To an observer standing on the surface of the earth, the celestial sphere appears to rotate about the celestial poles because of the earth's rotation. Objects fixed to the celestial sphere therefore appear to follow circular paths about the celestial pole. Some of these paths will cross an observer's local horizon twice a day (once for rising and once for setting). The exact circumstances of these *observed events* are a function of the associated object's celestial coordinates, the geographical latitude of the site and the local horizon profile at that site. The *declination* of the object determines *the points on the local horizon* at which rising and setting occur (if at all). The right ascension of the object then determines the times at which rising and setting occur.

Given that a particular declination path always intersects a given horizon profile at the same point, and that every horizon point is intersected by one and only one declination path, and always the same one, it follows that there is a unique, one-to-one mapping between points on a given horizon and an astronomical declination. This mapping will vary according to site latitude and horizon profile. So, given an alignment direction (azimuth), the elevation of the local horizon in the direction of the alignment and the geographical latitude of the site, we can calculate the astronomical declination of a (hypothetical) celestial body that would cross the visible horizon in the direction of the alignment. It is hypothetical because we don't yet know if a real celestial body actually moves along this declination path. At some point in our statistical assessments we can use such calculated declinations to see whether or not they coincide with the astronomical declination co-ordinate of any known celestial body.

Fixed stars (celestial objects whose apparent motion is so slow they do not appear to be moving in relation to us on earth) have constant declinations and cross the horizon profile at fixed alignments. However, planets, along with our Moon and Sun, vary their declinations, and therefore their alignments, over cycles of time. Nevertheless, as these changing declinations are a function of a planet's cycle they can be calculated for any particular 'time' and 'place'. For example, we can calculate what the declination of the Sun should be at its most northern and southern rising and setting points (summer and winter solstices) for any specified year. It is important to know that the Sun takes one year to complete its cycle where, in the northern hemisphere, a cycle is roughly equivalent to the Sun moving away from its most northern rising and setting points (summer solstice), rising further and further southward, until it is reaches its furthest point south along the horizon (winter solstice), finally returning to its most northern rising and setting points. The Moon, which has a more complicated form of movement, takes approximately 18.6 years to complete its (metonic) cycle.

For more detailed information on the astronomy used in this study see our *Supplementary Material.*

## 3. Archaeoastronomy and Standing Stones of Scotland

### 3.1. Bronze Age standing stones – previous work

In our re-analyses of Ruggles work, it was shown that megalithic monuments as a regional group were deliberately clustered in orientation towards (initially) unknown directions (Higginbottom and Clay 1999, S44, Tables 3 and 4, Higginbottom *et al.* 2001; Higginbottom *et al.* 2013: 25). These were either the internal alignments of monument elements, like the axis of the standing stones of a stone row or a thin, wide slab, or external alignments created by two monuments, such as a small stone circle and a standing stone, (see Ruggles 1984: Section 2 for details on site and alignment selection procedures). This was statistically confirmed for four out of six sub-regions: Uist, Argyll/Lorn, Mull/Coll/Tiree, and Islay/Jura (Higginbottom and Clay 1999, Higginbottom *et al.* 2001; Higginbottom *et al* 2013). Upon examination, Kintyre appears to have two separate orientation distributions, one more northerly and more southerly within the peninsular (Higginbottom 2003: 194). Lewis/Harris did not display any statistical support at that time but is now being re-examined due to new findings. Using comparative testing methods, but an entirely different test that was not as well suited to the circular data, Ruggles only found support for clustering on the isles of Islay/Jura. Other clustering was only found once the data were broken down into a number of archaeological classes and then re-tested (Ruggles 1984: 230-240).

Through further re-analyses of Ruggles work, in particular through our 2D & 3D landscape reconstruction, we have transformed our perception of the locational value of such sites by generating horizon profiles for sites in the above regions numerically and graphically from digital elevation data (Higginbottom *et al.* 2013). The 2D site horizon profiles contain information on distance, elevation and direction of the visible horizon viewed from the monuments and were used to test for the likelihood that the *distribution of observed horizon declinations* indicated by monument alignments for each region was consistent with a distribution of deliberate astronomical alignments rather than a distribution of random astronomical alignments (the latter being the null hypothesis). Thus digital data provided us with the means to create a *distribution of observed horizon declinations* and a random *distribution of expected horizon declinations* to carry out this test (explained below). These methods were made possible through the development of software by Smith, called *Horizon* (see **section 4**: **2D & 3D landscape reconstruction**).

Significantly, the digital elevation data allowed us to test the null hypothesis in a more rigorous way than Ruggles was able, due to improved computing power and accessible digital elevation data. Specifically, substantial computing power is required, plus good elevation data over an area of approximately 100 by 100 km, for determining the visible horizon profile of each site, something Ruggles did not have access to at the time. Ruggles attempted to overcome this problem and the various variable dependencies (Ruggles 1984: 254-261) with a number of statistical approaches in his 1980s and 1990s publications. In a similar way to the Thoms' work in 1978, 1980 and 1981, Ruggles created his expected data (null) from his observed data and then tested his original observed against this null expectation (Higginbottom 2003: 63-66). Explicitly, the approach used by Ruggles to examine his 276 observed orientations, was to create three lists: latitude of the monument, elevation of the horizon in the direction of the alignment and the possible range in degrees of each indicated azimuth (eg a possible viewing window of 0.1 … 1, 2 3 degree(s) or more depending upon alignment irregularities *etc.*), two of these lists were shuffled randomly and recombined with the first entry from each of the three lists "so as to produce 276 new latitude-altitude-width triplets.' Ruggles next steps were:

> (t)o each triplet we now assign an azimuth at random, and calculate the limiting declinations of the simulated indications. A set of 276 declination intervals produced in this way will satisfy the requirement that orientations have been randomised (1984: 255), as the property he wishes to randomise is structure orientation, and the property he wishes to investigate is the indicated declination (1984: 254)). He did this 100 times to create a probability statistic. He then compared the number of times the observed declination windows from the site orientations hit a particular declination value within a given declination range (up to $20^{o}$) compared with the number of times the expected random declination windows hit the same declination value within a given declination range. Ruggles tested for declinations running from $^{+}39^{o}$ to $^{-}35^{o}$ degrees.. Specifically, he asked: "(i) In how many of the 100 simulations were there as many or more hits upon this target (than the observed)? And (ii) In how many of the 100 simulations were there as many or fewer hits upon this target?" In this way he created a probability statistic that that gives the likelihood the occurrence of the 'observed events' could result from 'chance'.

However, there is a particular issue with this approach when one recognises the importance of the actual horizon profile (despite being the best one could do at that time). This approach does not test one set of real or physical data (observed) against another set of real or physical data (randomized expected). That is, a *real* randomised situation has not been tested against. The fundamental problem is that the horizon elevation function (the relationship between azimuth and elevation, and therefore too the associated declination) *is real and fixed to a specific site;* it is not a probabilistic distribution. Thus, this approach leads to results that are unphysical. This would mean, too, that Ruggles could not actually test with due consideration given to the actual landscapes even though this was an attempted consideration. Ruggles was aware of the methodological constraints at the time.

Lastly, in relation to this issue, Ruggles and colleagues made further attempts to deal with the real landscape given the limitations on data collection/availability. A variety of possible approaches are discussed in 1991, 1992 and 1996. In 1991, he and his colleagues used 'control locations' and collected a greater amount of horizon data pertaining to each site (eg horizon prominence, distance and direction). Such data were used for two major approaches called 'Global Analysis' and 'Local Analysis'. In 1992, they further investigated horizon qualities in relation to an association between particular astronomical phenomena and horizon prominence. All approaches required much extra fieldwork including horizon surveying and geographical data on nearby locations. However, it is clear that the aim is no longer testing the likelihood of the astronomy *per se*, but the choice of the locations themselves, the testing of declinations had not yet been solved.

Due to the scope of this paper we cannot critique all the methodologies discussed and applied, however we would like to make a few points. The applications carried out for the 1991 study of north Mull, whilst interesting, make it difficult to make generalisations from the statistics as the work focused on a small number of sites (there were seven sites from north of the island and 22 real control locations used for comparison) and, whilst measurements were taken around the $360^{o}$ horizon, they were only taken at large intervals, particularly where ranges intersected. So whilst much more real horizon data was gathered and the lack of easily accessible digital elevation data again has an impact on our knowledge of both control and case-study locations. These shortcomings were acknowledged by the authors. Their findings concluded that the stated hypothesis of "no difference between the *distribution* of visibility between monuments and control points" was rejected for one horizon distance category at $p<0.05$ and two categories at $p<0.1$, namely: (i) less than 1 km then (ii) 3-5 km and (iii) greater than 5 km, respectively. The *seven* sites 'eshewed distant horizons to the NW and NNW (that is greater than 5 km)' and ' avoided nearby horizons in the S and WSW (less than 1 km; 1991: S68; whilst the statistical tests for 12 sites of SR and SP in mid-Argyll in the appendix were not as convincing, the trend of avoiding nearby horizons in the ESE and WSW is similar to found for north Mull)'. The 'Local Analysis' of 1991 was a focused study on the site of Glengorm and asked whether 'the mere location of sites in areas of good settlement potential can explain the horizon profiles observed'. However, in the end, only four of the nine variables used to define comparative control sites, were consistent across the eight control sites. So, whilst the methodological ideas and outcomes were interesting, the comparisons and conclusions cannot be considered to be firmly connected to the original question for this local analysis.

There are no such concerns regarding 1992 paper. However, whilst the collection of data was clearly intensive, the number of sites is again a limiting factor as well as the issue of lack of extensive terrain data; together these reduce the ability to interpret the findings across a larger number of monuments or make clear generalisations. In emphasising the intensity of data collection, Ruggles and Martlew propose future work with GIS (Ruggles and Martlew 1992: S11). In the meantime, their conclusions for the placement of seven (7) stone rows in mid-north and NE Mull were: (i) there is an existence of a prominent peak to the east of south in a position associated with the rising of the Moon close to its southernmost position at the major lunar standstill and (ii) a non-local horizon (distant) in the south and west of south (1992: S12). In addition, they state that 'though by no means proven' (due to the low number of sites (n=3)), sites may have incorporated a second southern peak, placed more easterly than (i) above, to incorporate the rising of the minor lunar standstill rising of the Moon. Further, they add as 'an obvious speculation' that maybe there was an additional prominent peak set up with the rising or setting of the summer solstice Sun (n=3).

In 1996, Ruggles and Medyckyj-Scott looked to multiple viewshed analysis (first discussed in 1993). For archaeoastronomy, they wished to use this as a deterministic method 'to try to identify those natural features and astronomical events that best explain the observed placing of the stone rows (1993: 132)'. The preliminary visual results which are a lead up to the use of multiple viewsheds appear to support item (i) of their 1992 work as a first foray into using GIS. They argue that GIS can offer similar, and possibly more expansive, results whilst allowing access to, and use of, all possible terrain data in the area instead of relying on massive amounts of 'legwork' and hand-digitising of elevation and astronomical data. The five (5) other SS sites in north Mull in this paper do not contain a Ben More connection, and neither to, therefore, the visible astronomical connection of this peak. It is upheld by Ruggles and Medyckyj-Scott, by referring to the 1992 results of three sites, that there are strong suggestions that other mountains may fulfil this role. The actual multiple-viewshed data discussed in the methodology section to be used to observe or test these ideas (the examination of every possible nominated peak) every single horizon around that site is lacking in the results section. Therefore, whilst we have a very interesting possibility of an approach using multiple viewsheds and the determination of astronomical potential outlined, this paper does not accurately or clearly identify the connection between mountain and astronomy in Mull except in one possible direction and for a single prominent peak. Further, technical advancements were still not at the point were we could have the visual clarity to 'see' where the events happened on the horizons of the actual landscape that surrounded the monuments. This is an essential element for understanding the placement of the monuments and how astronomy and landscape are knitted together at these sites.

It was to overcome all of these limits in the study of archaeoastronomy discussed above that Smith developed the *Horizon* software which we apply below (Smith 2013).

## 4. 2D & 3D landscape reconstruction

### 4.1. Testing the distribution of observed horizon declinations indicated by monument alignments (2D) – orientation foci

The *distribution of expected horizon declinations* with which to compare the observed distribution was calculated by sampling each site's 2D horizon profile at the uniform intervals of 0.1 degree in azimuth, extracting the corresponding elevation and horizon distance for each of these azimuth points. Thus we had a distribution where every direction and its associated declination had been included, that is a distribution without obvious preferences. The azimuth and the national grid reference (converted to latitude) of each site or alignment, as well as the altitude of each sampled horizon point, allowed for the calculation of declinations along the horizon profile. This process produced a declination data file for each site tested with 3600 declinations each. Once all this was done, every single declination file was then concatenated according to each geographical region (Mull, Argyll, Lewis/Harris, Uist, Kintrye and Islay,) to produce 6 ultimate horizon declination files (Higginbottom *et al.* 2001). This process allowed us to discover that for the islands of Mull, Coll and Tiree together with 2 sites from North Argyll, as well as Argyll with Lorn, and Islay with Jura, the *distribution of observed horizon declinations* indicated by monument alignments was unlikely to be due to chance factors (Kolmogorov-Smirnov test, rejection of the null hypothesis: Mull *p*=0.00817 (n=24 sites; n= 40 declinations), Argyll *p*=0.00593 (n=21, n= 44 declinations) and Islay *p*=0.00105 (n=23, n= 41 declinations); Higginbottom *et al.* 2000: 46). We therefore interpreted this outcome to mean that the monument alignments indicating particular declinations *along the horizon* were deliberately chosen by the builders of these monuments. At this point, it wasn't clear which these preferred declinations were and whether they had astronomical significance. The most we could say was, combining this outcome with the support for deliberate clustering of alignments discussed above, that, *within each region* there was an interest or preference in creating alignments towards the *same declinations.*

In order to investigate this, we had to see where in the declination profile the differences occurred, and then determine if these declinations aligned with astronomical phenomena. Binning the observed and expected (random) data into 5-degree bins for each region (eg $0^0$ - $5^0$, $5^0$ – $10^0$), we applied a simple probability test (*p*), to test whether or not the number within each observed bin differed from that found in the expected (null) bins. This tests the likelihood of any difference in number occurring by chance. Once this test was done, the statistically supported bins were studied to see if they overlapped with declinations of astronomical bodies or phenomena (Higginbottom *et al.* 2001; Higginbottom *et al.* 2003: 140-143). Importantly, the statistically supported ranges could indicate an avoidance of, or clustering within, a declination range.

These analyses of orientation data, and their concomitant declinations, statistically supported an interest in the Moon's rising and setting points most close to the major and minor standstills both in the southerly and northerly directions, as well as the Sun at the winter solstice and flanking the equinoxes (see *Supplementary Material* for explanations of these astronomical events). We should point out that whilst no statistical support was found for the Sun at the summer solstice by region, a small number of sites were oriented in this direction within two degrees (approximately nine orientations out of 276 right across western Scotland). However, it will be seen further down in this section, that the summer solstice becomes important in other ways.

### 4.2. Standing stone landscapes in Bronze Age western Scotland (3D)

These above findings, along with the examination of our 3D landscape reconstructions of each site, revealed that the sky and land are woven together to create a complex series of particular interactions at very particular times of the lunar and solar cycles (Higginbottom et al. 2013).

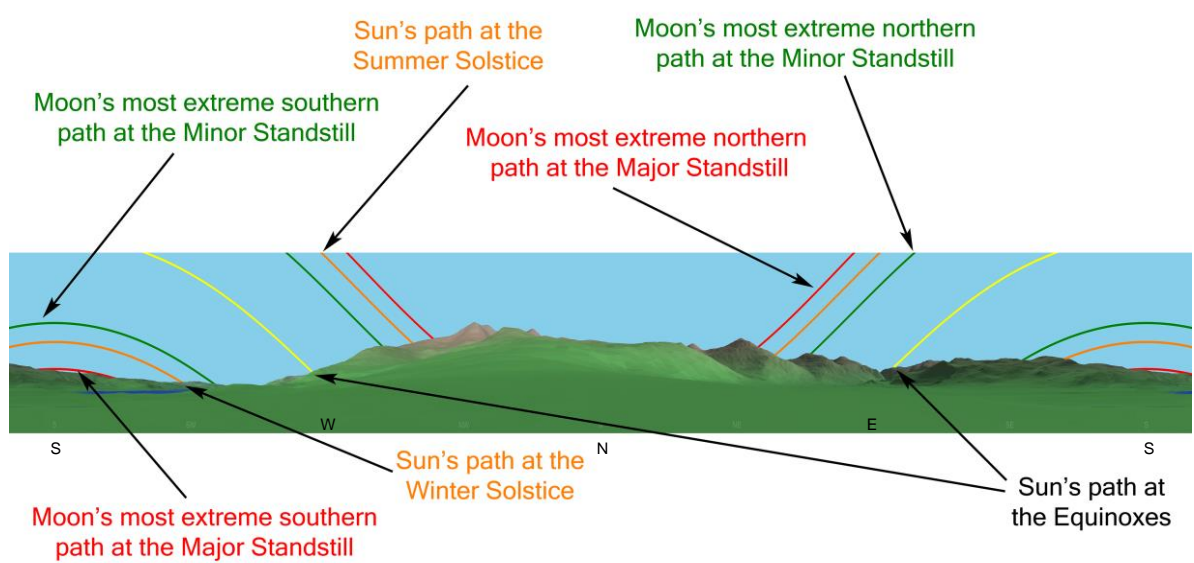

**Fig. 3: Example of 3-D rendering of the landscape** (classic site of Uluvalt on Mull) along with key to reading the paths of the sun and the moon on the other such figures below. N=north, S=south. Software created by Andrew Smith. Based upon the Ordnance Survey 1:50 000 Landform PANORAMA map with permission of the Controller of her Majesty's Stationery Office © Crown Copyright. For colour reproduction on the Web and in print.

Most importantly, we found two horizon landscape patterns, one that was basically the topographical reverse of the other. One or the other surrounded every site. For Coll and Tiree (n=6/6), and the majority of sites on Mull (n=9/16) there is a combination of *usual* visual cues, whether the sites are linear, single slabs, or small circular settings (Higginbottom *et al.* 2013: 47-53; Higginbottom *in preparation* 'The world begins here, the world ends here: Mull' (GH3)). We called these classic sites as they contained the first pattern we recognised, and the usual dominant cues for classic sites are (Fig. 4a-d & see Inline Supplementary Material (ISM) Figs. 1-4):

1. water is usually in the south (eg Fig 3, 4a & ISMFig. 1);
2. the northern horizon is closest, the southern most distant;
3. the northern horizon has a higher general profile or the highest vertical extents in the profile; the southern horizon has a very distinct dip (concave) or a lower general profile than the northern;
4. the highest areas of the northern and southern horizons focus around the four ordinal directions of NW, NE, SW and SE; occasionally the highest area is more northern if a single mountain or range fills the northern horizon (eg Fig 3, 4a & ISMFig. 1);
5. the highest points of the horizon profiles are usually made up of distinct mountains or hills; where there is no mountain or hill range, a single hill or higher ground is usually located near or at these compass points. Whilst most sites have relative peaks near all four ordinal points, some have three (e.g. Hough: NW, SW and SE, Fig 4c & ISMFig. 3);
6. the summer and winter solstitial Sun and standstill Moon tend to rise out of and set into these high ranges, hills, or ground.
7. a site most often forms an alignment internally, or with another site, at a lunar or solar orientation (the majority of which fall within the statistically supported declination ranges). For the Moon this is the Major or Minor Lunar Standstill (LS), and for the Sun it is at the winter or summer solstice (WS, SS). A few are aligned N-S; no sites are aligned near or on the midpoint between the solstices or equinox).

As mentioned above, those sites that do not reveal the landscape pattern above reveal a combination of reverse landscape traits, namely (see Fig. 4e & ISMFig. 5):

1. water is usually seen in the north;
2. the southern horizon is closest, the northern most distant;
3. southern horizon has the highest point(s) in profile; the northern horizon has a very distinct dip or overall lower horizon profile than the southern.

We call these simply 'reverse sites'. Their remaining astro-horizon qualities remain the same as those found in points 4-7 above, except that Argyll does have one alignment focused within a few degrees of the equinox (though obviously not a statistically supported event). For the sites investigated so far, we now know that Argyll (with Lorn; (n= 20) has roughly an equal number of each of the two astronomical-landscape series (GH & AGK Smith, i*n preparation, Intricate vistas: value driven landscapes in prehistoric Argyll* (GH4)). With these astro-landscape patterns, the occasional summer solstice alignment and the particulars found in the Supplementary Material (that a full moon at the major standstill in the south – the direction the majority of statistically supported orientations face - can only occur around the time of a summer solstice) shows us that the event of the summer solstice is likely just as firmly entrenched in the consideration of monument placement as those of the statistically indicated alignments.

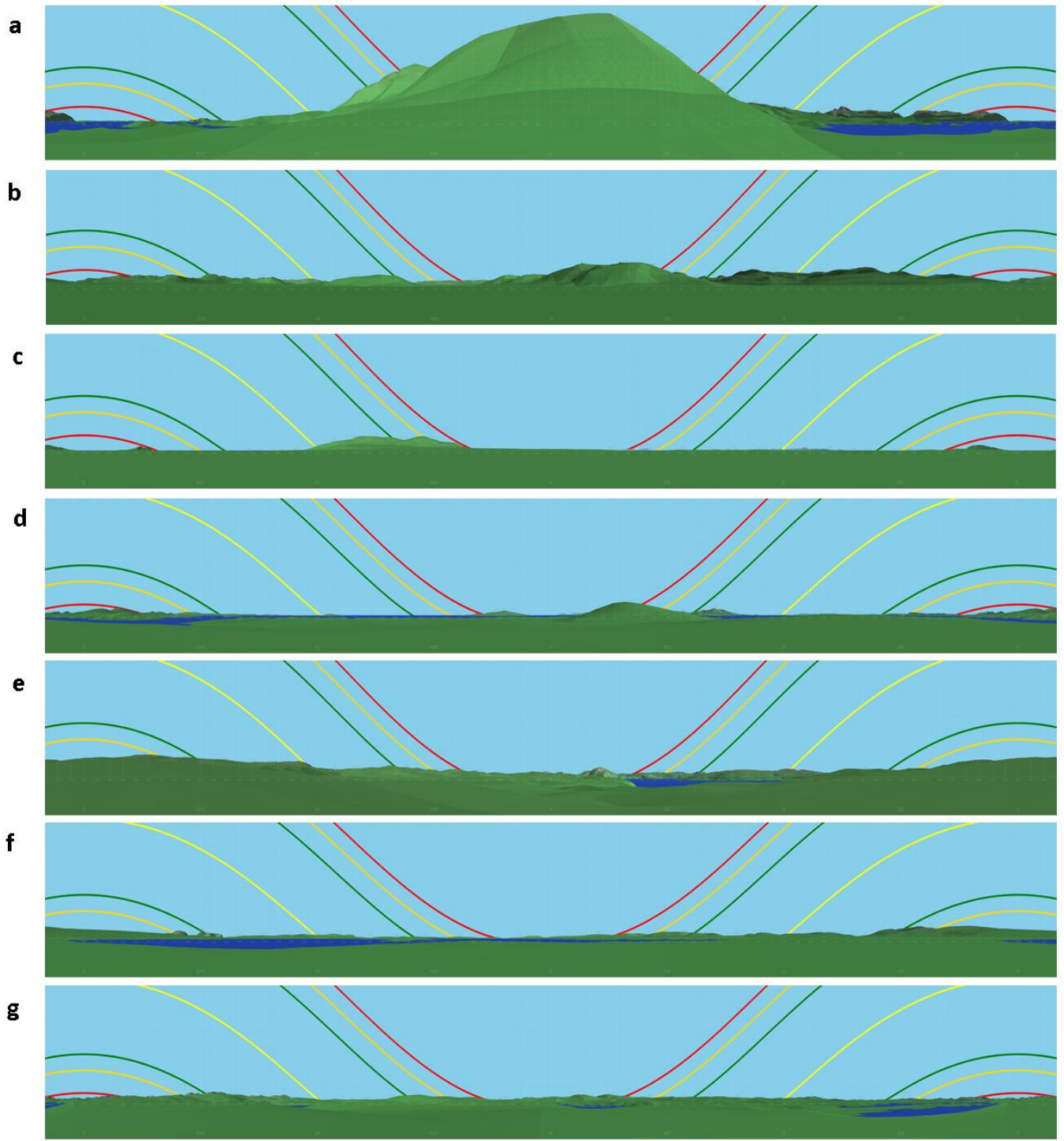

**Fig. 4. 3D landscapes of the Bronze Age example sites and the Neolithic great circles (See ISMFigs 1-5 & 8-9 for higher resolution and greater detail).** Note there is landscape overlap in the S for easier viewing. **a**, Cragaig, Mull (national grid reference (NGR) - NM4028 3901): a pair standing stones, approximately 4 m apart. One is a 1.3 m tall, the other is 1.6 m. **b**, Dunamuck, Argyll (AR84709290). Stone row of 4.4 m made up of three menhirs up to 3m high. **c**, Hough, Tiree: 2 small stone circles 90 metres apart and 40m in diameter. **d**, Blashavel, Uist (NF9122 8068): a thin, tall slab 2.5m high by 2m wide. Our pilot study on Uist has begun to find the same patterns as discovered elsewhere to date. **e**, Balliscate, Mull: a stone row of 3 irregular menhirs: 5 m in length and 1.8m to 2.8m in height. This is a reverse site. **f**, The Neolithic Stenness stone circle, on Orkney, is a reverse site. **g**, Neolithic Callanish I stone circle, Lewis. A.G.K. Smith, created all landscapes with the software Horizon, © A.G.K. Smith. Also created with Terrain 50. Contains Ordnance Survey data Crown copyright and database right (2012). HYPERLINK "http://www.ordnancesurvey.co.uk/docs/licenses/os-opendata-licence.pdf" http://www.ordnancesurvey.co.uk/docs/licenses/os-opendata-licence.pdf. For information on the 3D Horizon program settings for atmospheric clarity *etc.*, (see Supplementary Material). **For colour reproduction on the Web and in print.**

## 5.    The First Great Circles of Scotland: Callanish and Stenness

We have two distinct forms of evidence that support a connection between the Bronze Age standing stones just reviewed and the very first free-standing stone monuments of Scotland, the Great Circles, which arose in the late Neolithic: a great consistency between the orientation foci and the astro-landscape patterns.

To empirically assess the possible astronomical associations of Callanish and Stenness, we carried out two newly designed methods to formally *test the likelihood* of a connection between stone circles and astronomy. Ruggles in his major research project in Scotland chose to dismiss "from further consideration any on-site indications involving stone rings (1984: 61)".  This was because astronomical hypotheses involving sightings across stone rings are dependent upon other variables, such as whether or not the "site fits a particular geometrical construction". For instance, the need to know the shape of the monument in order to define the appropriate axes of the monument, which would in turn allow the determination of a central point from, or through, which an alignment might run. Further, no statistical test had yet been determined to deal with the associated probability issues connected to investigations of looking at orientations within a single circle. Such a test would involve separate determinations of the likelihood of various statistical errors, including errors in orientation due to archaeological alignment uncertainties and the uncertainty of which part of the astronomical phenomenon was of interest as it crossed the horizon (e.g. when it first touches the horizon or its final disappearance; thus testing the intentions of the builders).

In relation to Ruggles first concern, the possible and testable sightings for Callanish and Stenness were not deemed problematic. Both had been excavated and surveyed, and all the stone locations (missing and not) were accounted for, also therefore, the general shapes could be determined. Further, both of the monuments defined their own central focus via their internal structures: for Callanish it is the central standing slab and for Stenness it is the stone hearth. Their axes were defined in the same general manner, though the details differed due to their differing constructions. The axis of Callanish was determined by measuring the central line along the orientation of the large thin, central slab, which measured exactly N-S. Stenness' was ascertained by: (i) finding the centre point of the square stone hearth and the central point of the entrance of the monument, the one entrance-causeway, and running an extended line between the two. The Stenness axis-line runs $2^0$ west of N-S.

Another very important issue connected to circles with large numbers of outer stones, is the increased likelihood of hitting an astronomical object by chance, increasing the statistical errors, and therefore reducing the level of probability at which one can reject the null hypothesis. The number of stones for Stenness and Callanish are not high enough to increase the statistical errors significantly, and further, we have created a test that can take account of the relevant statistical errors related to alignment uncertainties and the uncertainty of which part of the astronomical phenomenon was of interest as the horizon was crossed.

## 5.1. The orientation foci of Callanish and Stenness

Looking along the entire 360 degree horizon of Figure 3 or 4a (ISMFig. 1) once again, it can be seen there are 8 possible extreme rising and setting LS targets and 4 extreme solstitial (looking at the coloured lines). Callanish 1 has 13 stones making up the circle and 3 close outliers (within 3.3-6.7 m, which are not part of the external linear stone rows). Altogether, the orientations of the large flat central slab (N-S axial alignment) plus those created from the alignments of this central stone to all the stones, contain 5 LS targets out of the possible 8 and 3 out of 4 possible solstitial targets (total = 8/12), along with the north and south cardinal points (Table 1 & ISMTable 1). Of these, all the risings and settings of the Major LS in the north and the south are accounted for (4/4), to which the majority of western Bronze Age SS sites on Mull, Coll, Tiree and Argyll are also aligned (Higginbottom et al. 2000). Interestingly, Mull, Coll and Tiree together showed a statistical preference for the Major LS both in the southerly and northerly directions ($p= 0.025$ and $p=0.077$, respectively) and Argyll the northern Major standstill ($p=0.026$) as well as the winter solstice ($p = 0.062$). Stenness, with its 12 stones and monument axis (the entrance with the central stone setting which creates a north/south cardinal axis contains 4/8 significant LS targets and two solstitial targets on opposite sides of the circle ($n=2/4$): the rising Sun at the SS and the setting at the WS. Together, Stenness and Callanish contain 7/8 LS targets and 3/4 solstitial targets. Notably, the Bronze Age sites as a group on Mull contain 7/8 possible rising and setting LS targets (GH3 *in preparation*) and Argyll contains all possible rising and setting LS targets (8/8); both regions have all solstitial targets (4/4) (GH4 *in preparation*). What we must determine now is the likelihood of these Callanish and Stenness results being due to chance.

## 5.2. Probability analyses of the orientations

For this *we devised* cross-correlation tests which compared the stone directions with the direction of the astronomical phenomena where it crosses the horizon. Specifically, we devised two tests that people can use. These assess the same basic enquiry, and though one is a more conservative version of the other, both tests incorporate the same appropriate errors in the construction of the data to be tested. Specifically, the less conservative test determines the probability that the overall monument *is not designed with astronomical considerations* (Test 1) and the more conservative test determines the probability that the overall monument *is designed with astronomical considerations* (Test 2).

As a *first step*, we made the assumption that the stone settings and alignments were fully determined and 'surveyed in' with posts, or similar, before any of the megaliths were put in place and that the most important requirement was that the stones then be placed centrally over the pre-determined aligned positions, rather than allowing a view of the exact alignment over or alongside the stones once the monument was set in place. We wished to test the assumption, then, that the most relevant goal was for the megaliths, the landscape, and the astronomy to be tied closely together rather than sighting the rising and setting from the completed monument on a particular day (this doesn't mean that the latter was not a second consideration). As we were only interested in the extreme rising and setting points at this stage, as supported by the statistical analyses to date (Ruggles 1984, 1985; Burl 2000; Higginbottom *et al.* 2013; GH3 *in preparation*; Patrick & Freeman 1988: 256-57), we excluded the testing of all astronomical and megalithic orientations related to, or close to, the cardinal points of east and west.

In carrying out the tests we: (i) calculated the azimuth range where the path of each astronomical body touched the visible horizon during its rising and setting (e.g. the Moon can skim along the horizon for several degrees before finally disappearing when setting). We called this range 'astronomical error' (Figs. 5-6 & ISMFigs. 6-7; Table 1); (ii) determined the orientations of the megaliths and the apparent axis of the circle; (iii) considered a possible reasonable, but strict, error that might be allowed for the orientations of the megaliths. Three degrees (+/-3$^o$) was chosen for Callanish and 3.5$^o$ for Stenness, for an alignment based securely within the parameters of each megalith's vertical edges, where the width of the megalith ranged from 5$^o$ to 10$^o$ as measured from the centre of each circle. This is the 'orientation error'; (iv) noted the number of megaliths that 'hit' the astronomical targets within 'total error' ('astronomical error' + 'orientation error') for each great circle (GC; observed hits; (Table 1)); (v) randomly selected circles of orientations by systematically and successively shifting the 'True North' points of each GC to the east by 0.5$^o$. Seven hundred and nineteen (719) shifted circles (ShCs) resulted for each GC from these 0.5$^o$ steps. We could then calculate and graph (vi -a) the distribution of the number of hits on the astronomical targets for the 719 ShCs within total error (Figs. 5-6 & ISMFigs. 6-7) and observe (vi-b) the number of ShC hits (expected on a randomised basis) compared to the GC data.

### 5.3. Test One: simple assessment. Hypothesis one - the monument is not designed with astronomical considerations

Using our original data (observed data), we compared it to the data from each shifted circle (n=719). We did this by dividing the outcome of (vi-b) (the number of ShCs which hit the same number of targets, or more than, the GC) by the total of number of the series of ShCs (719 + theoriginal GC; *n*=720). For Callanish, vi-b is nine (9) and for Stenness twenty-seven (27) and the concomitant probabilities are *p*=0.0125 and *p*=0.0166, respectively. Thus the likelihood of the number of 'hits' coming from random chance is 1.25% and 1.66%, or *more specifically the probability that the circles are not astronomical is low* (see Figs. 5 & 6).

### 5.4. Test Two: conservative assessment of a monument's potential astronomical design. Hypothesis 2 - the monument is designed with astronomical considerations

For this test, we smooth the data by +/-1.5$^o$ to reduce the general noise and we judge support for an astronomical design of the monument by finding the number of 'hits' for orientations close to zero shift to be substantially above the general level of those found in the 719 ShCs. On smoothing the data we found that Stenness had its highest peak of hits (out of any monument orientation) at a level of 6, with a shifted orientation of two degrees 2$^o$ west and Callanish had its second largest peak of hits at a level of 7 for an orientation shifted 4° to the east (Figs. 5b & 6b, respectively). We can now take the peaks of the independent samples and compare them with the peak that *covers zero*, which is the peak of the GC itself; this actual peak may maximize to either side of zero itself as noted above. Such peak shift shows a maximizing away from zero which may reflect an error in the orientation of the stones, for example. An 'independent sample' is where a peak or curve has no overlaps with another and its width can be calculated using an approximation of the error range appropriate to our measurements of the original GC. Therefore, this error range was taken to be related to an average of the 'total error' used to create the distributions, viz: 7.64° for Stenness and 7.67° for Callanish. With an orientation uncertainty of +/-3° and an astronomical uncertainty for rising and setting up to the same order we find a total of approximately 47 independent samples associated with a peak (specifically 47.1 for Stenness and 46.9 for Callanish). This was found by dividing the 360° by the average total error range.

|  | Megalith number from plan | Orientation of megalith from central slab – to vertical centre of stone | Astronomical phenomenon range (astronomical error) | Possible astronomical phenomenon (target) | Target 'hit' (within orientation error) |
|---|---|---|---|---|---|
| **Callanish** | | In degrees | In degrees | | |
| central slab axis* | 29 | 0 | 0 | 'True North' | |
| | 53 | 5 | | | |
| | 41 | 27 | 28.2-30 | MajLS rise (nth) | Yes |
| close outlier | 34 | 40 | 39.6-41.5 | SS rise (nth) | Yes |
| | 42 | 52 | 55-56 | MinLS rise (nth) | Yes |
| | 43 | 77 | | | |
| | 44 | 98 | | | |
| | 45 | 122 | 128.9-130 | MinLS rise (sth) | |
| | 46 | 142.5 | 140.1-141.4 | WS rise (sth) | Yes |
| close outlier | 35 | 162 | 163.5-166 | MajLS rise (sth) | Yes |
| central slab axis | 29 | 180 | 180 | south | |
| | 47 | 183 | | | |
| close outlier | 9 | 196 | 188.2-194.2 | MajLS rise (sth) | Yes |
| | 48 | 214 | 215-216.5 | WS set (sth) | Yes |
| | | | 226.2-227 | MinLS set (sth) | |
| | | | 303.2-304.4 | | |
| | 49 | 253 | | | |
| | 50 | 292 | | | |
| | | | 317.9-319.2 | SS set (nth) | |
| | 51 | 328 | 331-334.3 | MinLS set (nth) | Yes |
| | 52 | 349.5 | | | |
| **Stenness** | | In degrees | In degrees | | |
| axis of entrance + hearth | | 2 | 0 | 'True North' | |
| | 8 | 10.5 | | | |
| | | | 24.8-27 | MajLS rise (nth) | |
| | 9 | 42 | 40-42 | SS rise (nth) | Yes |
| | | | 55-56 | Min LS (nth) | |
| | 10 | 74 | | | |
| | 11 | 104.5/105 | | | |
| | 12 | 138 | 136-137.6 | MinLS rise (sth) | Yes |
| | | | 147.2-151 | WS rise (sth) | |
| | 1 | 162 | | | |
| | | | 173 | MajLS rise -glimmer | |
| axis of entrance + hearth | | 178 | 180 | south | |
| | 2 | 191 | 192 | MajLS set -glimmer ends | Yes |
| | 3 | 218 | 213.8-214.7 | WS set (sth) | Yes (3.3) |
| | | | 225-226.5 | MinLS set (sth) | |
| | 4 | 251 | | | |
| | 5 | 282 | | | |
| | 6 | 310 | 306.1-307.5 | MinLS set (nth) | Yes |
| | | | 321.5-323 | SS set (nth) | |
| | 7 | 339.5 | 336-338.8 | MajLS set (nth) | Yes |

**Table 1 (See SMITable 1):** Observed data used for probability calculations testing the astronomical associations of the great circles, Callanish and Stenness. Note that whilst Stone 3 of Stenness was 0.3 degrees outside of the orientation error we included it in our assessment. There are no other orientations this close outside of the +/-3 degree orientation error range. **For colour reproduction on the Web**

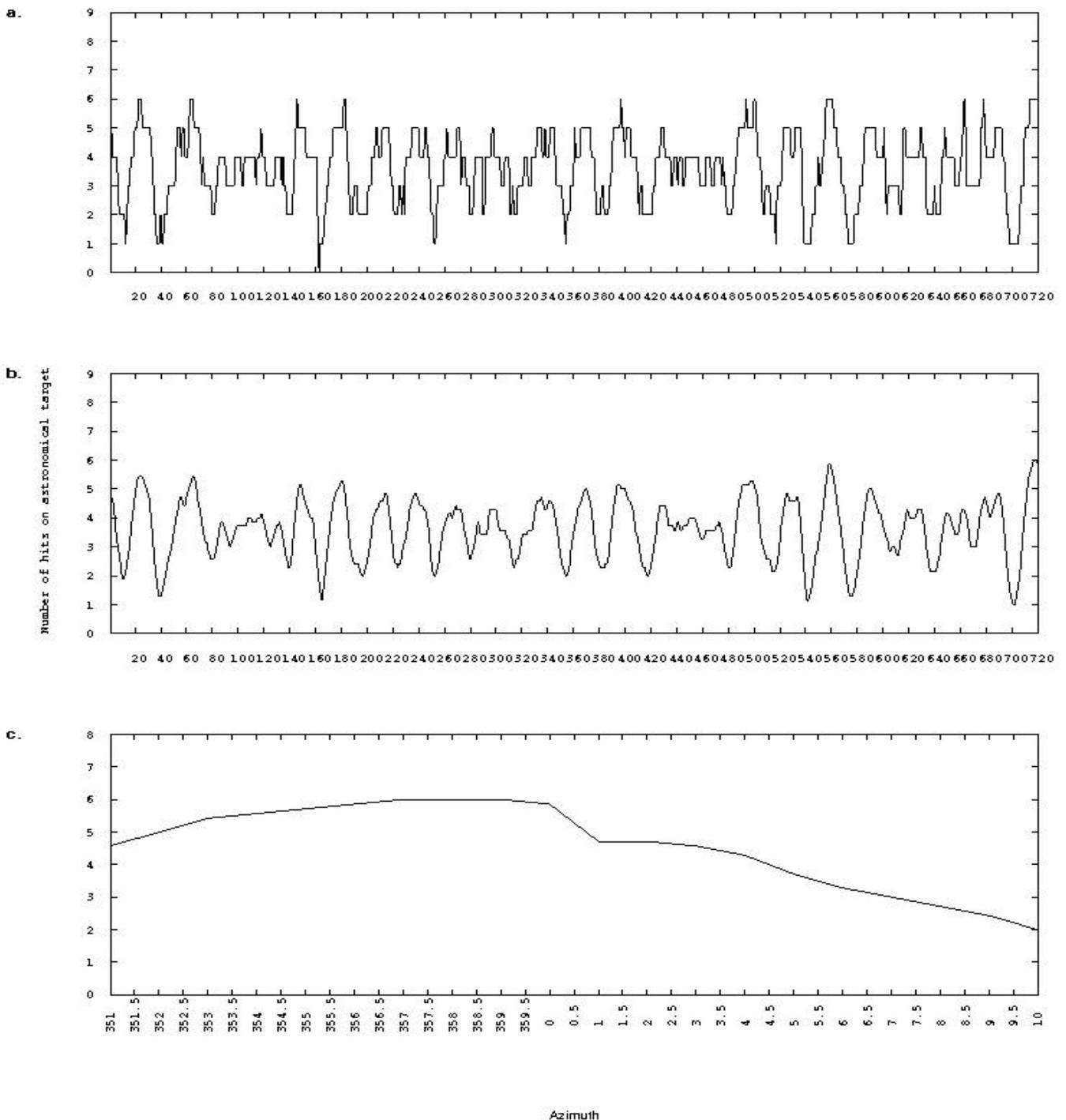

**Fig. 5 (SMFig. 6) Expected data of the shifted circles used for Stenness' probability calculations.** These data were used to test the astronomical associations of this great circle. Zero (0°) on the x- axis represents "True North'. As a whole the graph indicates the number of hits on the astronomical targets for each shifted circle in relation to the location of that shifted circle's north point, shown the x- axis. There is one shifted circle every 0.5°. Thus, the y-axis indicates the number of hits on the astronomical targets for each of the 719 shifted circle and the x-axis indicates the azimuth or orientation. Section 5a, is the original curve, 5b is the smoothed data and 5c shows the shape and height of the peak closest to zero shift for the smoothed data. Note that for Stenness whilst Stone 3 was 0.3° outside of the +/- 3.0° orientation error we chose to include it in our assessment, to do so meant increasing our orientation error by 0.5° for the entire statistical assessment of Stenness. There are no other orientations this close outside of the +/-3 degree orientation error range, in fact, apart from the same stone being within 3.0° of total error of an alternative phenomena the remaining five astronomical phenomena do not have any possible stone alignments at all, as defined by us, as can be seen from Table 1.

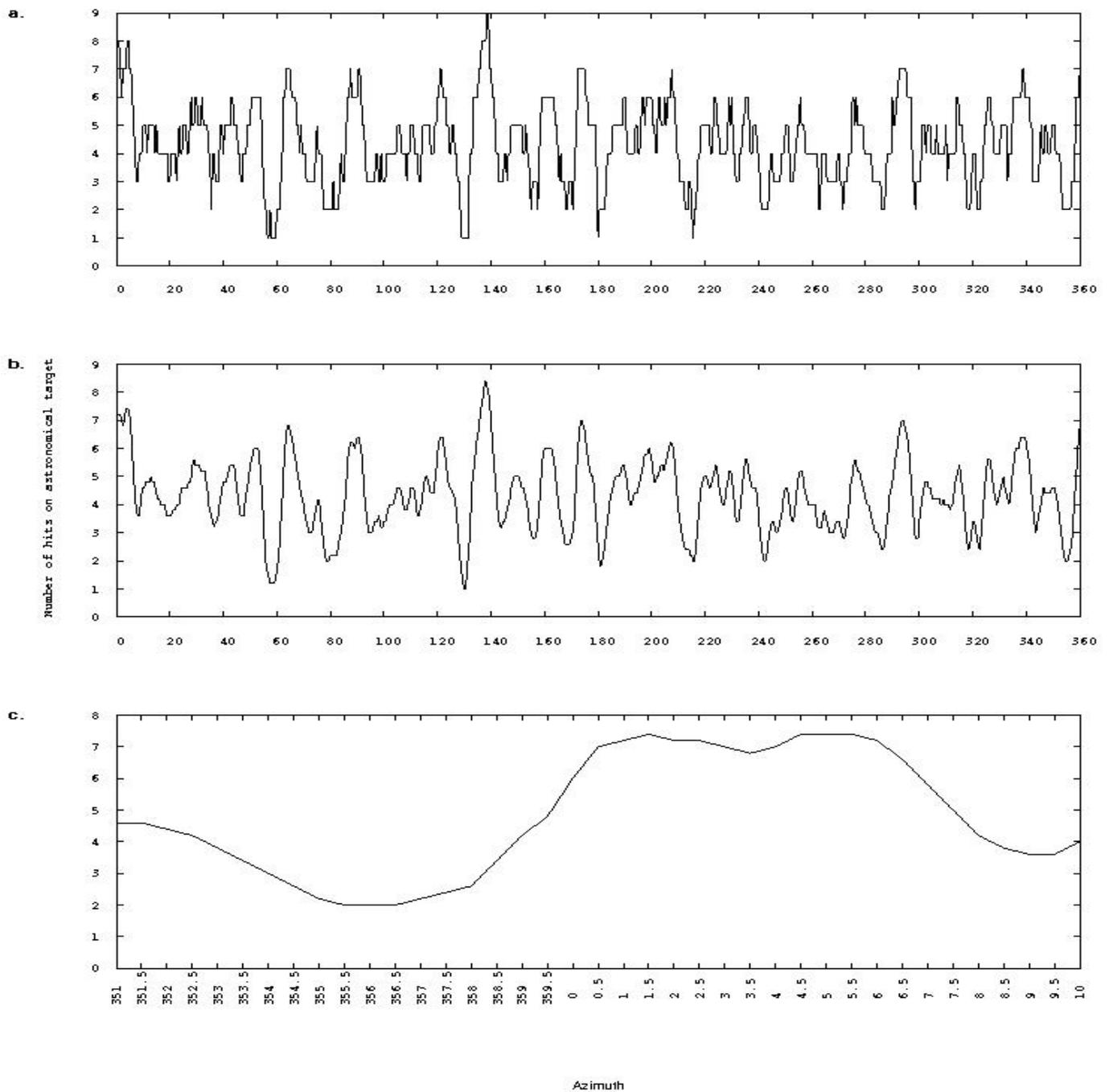

**Fig. 6 (ISMFig. 7): Expected data of the shifted circles used for Callanish's probability calculations**. These data were used to test the astronomical associations of this great circle. Zero (0°) on the x- axis represents "True North'. As a whole the graph indicates the number of hits on the astronomical targets for each shifted circle in relation to the location of that shifted circle's north point, shown on the x- axis. There is one shifted circle every 0.5°. Thus, the y-axis indicates the number of hits on the astronomical targets for each of the 719 shifted circle and the x-axis indicates the azimuth or orientation. Section 6a, is the original curve, 6b is the smoothed data and 6c shows the shape and height of the peak closest to zero shift for the smoothed data.

Our probability calculation with these numbers to test whether monuments *were designed with astronomical considerations in mind* is: 1- ((vi-b) / (*n* of independent samples)): Callanish: 1 – (2/46.9). As no independent sample has more hits than Stenness, we calculate the probability to be > 1-(1/47). Therefore the probability of Stenness *being designed with astronomical considerations* is *p* > 0.979 and that of Callanish is *p* = 0.957 (where 1=true and zero=not true). Or to put it another way, the likelihood of the monuments being astronomical is: 97.9% for Stenness and 95.6% for Callanish.

### 5.5. 3D landscape reconstructions of Callanish and Stenness

Applying our 3D landscape models to the great circles of Callanish and Stenness, we find that these shared a combination of the astronomical and landscape cues found at the Bronze Age sites more than 1500 years later, where Callanish on the western Isle of Lewis is a classic site and Stenness in the north on Orkney, a reverse site. Despite Stenness having an almost flat horizon we see that the builders engineered very particular horizon views (Fig. 4f; & ISMFig. 8). The northern cardinal point is closely marked with a horizon notch, with the rising and setting Major Moon at the LS placed very close to 25 degrees either side of this. The SS Sun and northern Major Moon at the LS both rise out of a northern slope of the high ranges in the NE and set into the high points in the NW; the Minor Moon at the LS in the north rises out of the top of a hill in the highest range in the NW. The WS Sun rises out of the closest and highest range in the SE and sets in one of two very distinctive ranges in the SW; the Minor Moon at the LS in the south does the same, setting into the only other significant peak in the SW. Also in the south, the 'top' rim of the Moon of the Major LS lies just below the horizon within 0.5 degrees in declination, and travels along this declination below the horizon, from 178 degrees to 187 degrees. What is important, here, is that its glow could be seen travelling above the horizon for nearly 10°. The equinox Sun rises out of, and sets into, ranges east and west of the site.

For Callanish, the highest points in the distant north sit NW and NE, as expected (Fig. 4g; & ISMFig. 9). The Major Moon there rises out of the slopes of the NE range at the LS and begins to set into one of the peaks in the NW before rolling down its slope. Both the SS Sun and Minor Moon at the LS rise out of an undulating horizon, with the SS Sun again setting into the highest point in the NW. In the south the Major Moon rises out of the closest and highest horizon in the SSE and sets into ranges in the SSW. The WS Sun and the Minor Moon set in one of two very distinctive ranges in the SW. From the examination of these late Neolithic landscapes of the GC (Fig. 4f-g), it is clear that we can see a strong resemblance with those from the Bronze Age (Fig. 4a-e). Their landscape and astronomical choices in relation to horizon distances and direction, horizon profile (mountain/hill locations, lower and higher ground) and the astronomical phenomena associated with each of these - are clearly consistent, so too are the monument alignments (as demonstrated above).

## 6. Discussion

From the beginning our *Western Scotland Megalithic Landscape Project* has innovatively applied approaches to data analysis which are used in mainstream astrophysics, with which some of us have previously been involved. Such approaches have involved the devising and development of new statistical tests, the selection of tests from a broader suite than commonly accessed in archaeoastronomy, and rigorous, computer-intensive, determination of confidence limits when the tests are applied to archaeoastronomy situations. This has resulted in the use of tests that are more appropriate for the data being tested (e.g. Higginbottom and Clay 1999), leading to well-defined and convincing advances in the study of megalithic astronomy in the British Isles. The present probability analyses are no exception.

For the first time statistical tests have been constructed to test the astronomical potential of *single* standing stone circles which have been the bane of many an archaeoastronomer due to the large amount of potential random errors and background noise as well as the inherent number of statistical trials. The latter can be a particular worry. Specifically, the more stones one has in a circle, the greater the numbers of trials (the testing of each stone alignment is a trial) which leads to higher *p (*probability) values. This means, one is increasing the probability (or likelihood) that the pattern one has observed is due to chance. Therefore, one needs tests that can factor in all these trials, and we have this in our version of a cross-correlation test. Therefore, in the present work, we have tested for significant structural design properties in standing stone circles and then tested those properties against the possibility that they may be astronomically-related, taking into account the actual horizon viewing structure at individual sites. This work has involved care in assessing, and minimising, statistical penalties involved with the selection of hypotheses, and the due consideration of the effect of large numbers of possible alignments in cases when many stones are found in a circle. The application of these tests has created a break-through in the quest for when and where the use of complex astronomical and landscape patterns were *first* associated with standing-stone structures in Scotland, possibly all of Britain. The statistical results for Stenness and Callanish are compelling.

We deduce that whilst there is a diversity of archaeological expression in site architecture of the case-study sites we examined through the Neolithic and Bronze Ages (SStS, SC, SR, stone pairs) as well as varied detailed monument associations (number and kinds of monuments found close together or in sight of one another, including mounds, cairn and cist varieties and burial styles: Armit 1976, Richards (a) in Richards 2013) and possible site activities (burning events, local ritual ploughing: Ashmore 2002; Ritchie 1976; Richards and Wright in Richards 2013;, Richards (b) in Richards 2013) there are clearly shared, abiding values to be discovered. Through the combination of an archaeological review (Higginbottom *et al.* 2013), appropriate supporting statistical tests (here as well as, for example, 1999, 2000) and the examination of many individual 3D astro-landscapes (here as well as 2013, 2001, 2000, Higginbottom 2003) we have experienced and understood something of the general and the localised concerning free-standing stones. In relation to astronomy, we have the repeated use of an interest in solar and lunar extreme risings and settings along the horizon. Importantly, there are also combinations of astronomical targets at many Bronze Age sites just as at the Great Circles. For example, opposite directions of a single stone row, or a combination of an internal alignment and an alignment with another site, can contain two different lunar alignments; or two parallel monuments side-by-side or close-by might contain a solar and one or more lunar alignments (GH3 *in preparation*; Ruggles 1985*).* Relevantly, Bronze Age sites can cluster together or nearby such that a larger number of targets are covered within a small local region, such as in the Kilmartin Valley and in this way, perhaps, a small area may have a similar function to one of the past great circles. Further, there is a consistent association of these constructed locales with the dead, most usually cremated.

In relation to astro-landscape features only, it can be argued that the localised variations (like 3 prominent hills/ranges/elevations instead of 4 in the ordinal directions of NW, NE, SW and SE) could be put down to finding landscapes with as many of the key features as possible in their local area. So whilst "we observe the physical residues of a series of 'highly localised' social encounters of ritual (Barrett 1994, 72), they are clearly "organised within a framework of wider cultural motifs" (Duffy 2007: 54), and through their chosen astronomical emphasis within regions, "manifest as a distinct entity of local time, place and experience" (Duffy 2007: 54).

7. **Conclusion**

In conclusion, the local visual dominance of the first great circles in the north of Britain seems to have led a cultural transformation that connected standing stones to the local landscape and the orderly arrangement of the Universe across Scotland. Soon after these sites were created, a number of Late Neolithic stone monuments were erected especially circles and pairs or single standing stones (Higginbottom *et al.* 2013: 3-8) a process which continued until the Chalcolithic/Early Bronze Age. By the end of the Bronze Age (approximately 800 BC); hundreds of smaller stone circles and settings existed (Burl 1993, 2000; Higginbottom *et al.* 2013: 16-18) and it is these later monuments that continued the tradition of connecting with a cosmological-landscape ideal that was first set in standing stone more than 2000 years previously, demonstrating the longevity and relevance of this cosmological system.

**Acknowledgements**

Andrew GK Smith provided specialist advice on astronomy. AGKS designed and tested the 2D and 3D software (***Horizon)*** for the creation of the 2D and 3D landscape models and assisted with the design of the original empirical analyses. We thank P. Ashmore for his critical assessment of an earlier version of the paper and providing access to the phase tables of his forthcoming work on the excavations of Callanish. We also thank Douglas Scott for his survey data of the azimuthal measurements of the Stenness stone circle to compare with various historical archaeological plans and his photograph of Stenness.

## 8. References

Armit, I.. *The Archaeology of Skye and the Western Isles*. (Edinburgh University Press, Edinburgh, Edinburgh, 1996).

Ashmore, P. J. *Calanais Survey and Excavation 1979-88* with contributions by T Ballin, S Bohncke, A Fairweather, A Henshall, M Johnson, I Maté, A Sheridan, R Tipping and M Wade Evans. (*in press*).

Ashmore, P. *Calanais Standing Stones*. (Historic Scotland, (2002).

Ashmore, P. J. Radiocarbon dating: avoiding errors by avoiding mixed samples. *Antiquity* 73 (279): 124–130 (1999).

Barber, J. W. The excavation of the holed-stone at Ballymeanoch, Kilmartin, Argyll. *Proceedings of the Society of Antiquaries of Scotland,* 109: 104-11 (1977-78).

Burl, A. *From Carnac to Callanish: The Prehistoric Stone Rows and Avenues of Britain, Ireland, and Brittany* (Yale University Press, New Haven, 1993).

Burl, A. Pi in the sky. In D.C. Heggie (ed.), *Archaeoastronomy in the Old World*, 141-69. (Cambridge University Press, Cambridge, 1982).

Burl, A. *The Stone Circles of Britain, Ireland and Brittany*. (Yale University Press, New Haven, 2000).

Burl, A. *The Stone Circles of the British Isles* (Yale University Press: New Haven, 1976).

Downes, J. (ed) ScARF *Summary Bronze Age Panel Report*. (June 2012), Scottish Archaeological Research Framework: Society of Antiquaries of Scotland. Version 1, available online at HYPERLINK "http://tinyurl.com/clxgf5s"http://tinyurl.com/clxgf5s

Duffy, P. R. J. Excavations at Dunure Road, Ayrshire: A Bronze Age cist cemetery and standing stone. Proceedings of the Society of Antiquaries of Scotland, 137, 69–11 (2007). HYPERLINK http://archaeologydataservice.ac.uk/archiveDS/archiveDownload?t=arch-352-1/dissemination/pdf/vol_137/137_069_116.pdf

Higginbottom, G. *Interdisciplinary study of megalithic monuments in western Scotland* Unpublished PhD thesis. University of Adelaide,(2003).

Higginbottom, G. and Clay, R. Reassessment of sites in Northwest Scotland: a new statistical approach. *Archaeoastronomy* 24: S1-S6 (1999). HYPERLINK http://adsabs.harvard.edu/full/1999JHAS...30...41H

Higginbottom, G., Smith, A., Simpson, K. and Clay, R. Gazing at the horizon: heavenly phenomena and cultural preferences within northwest Scotland? In C. Esteban and J. Belmonte (ed) *Astronomy and Cultural Diversity*. Pp. 43-50. (Organismo Autónomo de Museos de Cabildo de Tenerife: Teneriffe, 2000).

Higginbottom, G., Smith, A., Simpson, K. and Clay, R. Incorporating the natural environment: investigating landscape and monument as sacred space. In M. Gojda and T. Darvill (eds.), *Landscape archaeology: new approaches to field methodology and analysis*. British Archaeological Reports S987: 97-104. Oxford: Archaeopress (2001).

Higginbottom, G. and Smith, A. and Simpson, K. and Clay, R. More than orientation: placing monuments to view the cosmic order. In A, Maravelia (ed), Ad Astra per Aspera et per Ludum: European Archaeoastronomy and the Orientation of Monuments in the Mediterranean Basin. British Archaeological Reports, International Series S1154: 39–52. Oxford: Archaeopress (2003).

Higginbottom, G., Smith, A.G.K and P. Tonner. A re-creation of visual engagement and the revelation of world views in Bronze Age Scotland in *Journal for Archaeological Theory and Method* (2013). HYPERLINK http://link.springer.com/article/10.1007/s10816-013-9182-7

Lynch, F. The Later Neolithic and Earlier Bronze Age. In F. Lynch. S.Alderhouse-Green. and J. Davies (eds.), Prehistoric Wales, 79-138. (Stroud: Sutton, 2000)

Martlew, R.D. and Ruggles, C.L.N. Ritual and landscape on the West Coast of Scotland: an investigation of the stone rows of Northern Mull. *Proceedings of the Prehistoric Society* 62: 125-129 (1996).

MacKie, E. The prehistoric solar calendar: an out of fashion idea revisited with new evidence. *Time and Mind*, 2.1 (March), 9-46 (2009).

Mullin, D. Remembering, forgetting and the invention of tradition: burial and natural places in the English Early Bronze Age. Antiquity 75, 533-7 (2001).

Owoc, M., The times, they are a changin': experiencing continuity and development in the Early Bronze Age funerary rituals of southwestern Britain. In J. Brück (ed.), Bronze Age landscapes: tradition and transformation, 193-206. (Oxford, Oxbow, 2001).

Ritchie, J.N.G. The Stones of Stenness, Orkney. *Proceedings of the Society of Antiquities of Scotland*, 107: 1-60 (1976).

Richards, C. (Editor) *Building the Great Stone Circles of the North* (Windgather Press, Oxford, 2013).

Richards, C (a) Interpreting Stone Circles in C. Richards (Editor) *Building the Great Stone Circles of the North* (Windgather Press, Oxford, 2013).

Richards, C (b) Wrapping the hearth: constructing house societies and the tall Stones of Stenness, Orkney in C. Richards (Editor) *Building the Great Stone Circles of the North* (Windgather Press, Oxford, 2013).

Richards, C. & Wright, J., Monuments in the making: the stone circles of Western Scotland in Richards, C. (Editor) *Building the Great Stone Circles of the North* (Windgather Press, Oxford, 2013).

Royal Commission of Ancient & Historical Monuments (RCAHMS) *CANMORE internet database.* Canmore ID 4156, Site Number NB23SW 1 which lists the Radio Carbon Dates along with their archaeological context:

HYPERLINK http://canmore.rcahms.gov.uk/en/site/4156/details/lewis+callanish/

Patrick J. and Freeman P. A cluster analysis of astronomical orientations in (C.L.N. Ruggles (ed) *Records in Stone.* Papers in memory of Alexander Thom, 251-261 (Cambridge, Cambridge University Press, 1985

Ruggles, C. L. N. Megalithic astronomy: A new archaeological and statistical study of 300 Western Scottish Sites. Oxford: British Archaeological Reports British Series 123 (1984).

Ruggles, C. The linear settings of Argyll and Mull. *Archaeoastronomy* (JHA) no. 9 (1985) S105–132. HPERLINK http://adsabs.harvard.edu/full/1985JHAS...16..105R

Ruggles, C.L.N. and Martlew, R. The North Mull Project 3: prominent hill summits and their astronomical potential. *Archaeoastronomy* 17: S1-13*. Journal for the History of Astronomy*, supplement to vol. 23 (1992). HYPERLINK http://adsabs.harvard.edu/abs/1992JHAS...23....1R

Ruggles, C.L.N. *Megalithic Astronomy: A New Archaeological and Statistical Study of 300 Western Scottish Sites* (British Archaeological Reports British Series 123: Oxford, 1984).

Ruggles, C.L.N., Martlew, R.D. and Hinge, P. The North Mull Project 2: the wider astronomical potential of the sites. *Archaeoastronomy* 16: S51-75. *Journal of the History of Astronomy*, supplement to vol. 22 (1991). HYPERLINK http://adsabs.harvard.edu/abs/1991JHAS...22...51R

Schulting, R., Sheridan, R., Crozier, R and Murphy, E. Revisiting Quanterness: new AMS dates and stable isotope data from an Orcadian chamber tomb. *Proceedings of the Society of Antiquaries Scotland* 140: 1–50 (2010). HYPERLINK http://archaeologydataservice.ac.uk/archiveDS/archiveDownload?t=arch-352-1/dissemination/pdf/vol_140/140_001_050.pdf

Sheridan, A. & Brophy, K. (eds) *ScARF Summary Neolithic Panel Document* (June 2012), Scottish Archaeological Research Framework: Society of Antiquaries of Scotland. Version 1, available online at HYPERLINK "http://tinyurl.com/d73xkvn"http://tinyurl.com/d73xkvn

Sheridan, J.A. The National Museums' Scotland radiocarbon dating programmes: results obtained during 2005/6. *Discovery and Excavation in Scotland* 7 (2006).

Smith, A.G.K., *Horizon User Guide and Implementation Notes. Documentation Version 0.12* December 3, 2013: 13 (http://www.agksmith.net/horizon).

Whittle, A., Healy, F., and Bayliss, A. Gathering Time: dating the early Neolithic enclosures of southern Britain and Ireland. (Oxbow Books, Oxford, 2011).

**Figures Captions for Inline Supplementary Material (For colour reproduction on the Web)**

**Figure ISM1** High resolution, 3D landscapes of Cragaig, Mull. Created with the software *Horizon* by A.G.K. Smith, © A.G.K. Smith. Created with Terrain 50. Contains Ordnance Survey data Crown copyright and database right (2012). http://www.ordnancesurvey.co.uk/docs/licenses/os-opendata-licence.pdf

**Figure ISM2** High resolution,3D landscapes of Dunamuck, Argyll. Created with the software *Horizon* by A.G.K. Smith, © A.G.K. Smith. Created with Terrain 50. Contains Ordnance Survey data Crown copyright and database right (2012). http://www.ordnancesurvey.co.uk/docs/licenses/os-opendata-licence.pdf

**Figure ISM3** High resolution, 3D landscapes of Hough, Tiree. Created with the software *Horizon* by A.G.K. Smith, © A.G.K. Smith. Created with Terrain 50. Contains Ordnance Survey data Crown copyright and database right (2012). http://www.ordnancesurvey.co.uk/docs/licenses/os-opendata-licence.pdf

**Figure ISM4** High resolution, 3D landscapes of Blashavel, Uist. Created with the software *Horizon* by A.G.K. Smith, © A.G.K. Smith. Created with Terrain 50. Contains Ordnance Survey data Crown copyright and database right (2012). http://www.ordnancesurvey.co.uk/docs/licenses/os-opendata-licence.pdf

**Figure ISM5** High resolution, 3D landscapes of Balliscate, Mull. Created with the software *Horizon* by A.G.K. Smith, © A.G.K. Smith. Created with Terrain 50. Contains Ordnance Survey data Crown copyright and database right (2012). http://www.ordnancesurvey.co.uk/docs/licenses/os-opendata-licence.pdf

**Figure ISM6** Expected data of shifted circles used for Stenness' probability calculations. These data were used to test the astronomical associations of this great circle. Zero ($0^o$) on the x- axis represents "True North'. As a whole the graph indicates the number of hits on the astronomical

targets for each ShC in relation to the location of that ShC's north point, shown the x- axis. There is one ShC every 0.5°. Thus, the y-axis indicates the number of hits on the astronomical targets for each of the 719 ShC and the x-axis indicates the azimuth or orientation.

**Figure ISM7** Expected data of shifted circles used for Callanish's probability calculations. These data were used to test the astronomical associations of this great circle. Zero (0°) on the x- axis represents "True North'. As a whole the graph indicates the number of hits on the astronomical targets for each ShC in relation to the location of that ShC's north point, shown the x- axis. There is one ShC every 0.5°. Thus, the y-axis indicates the number of hits on the astronomical targets for each of the 719 ShC and the x-axis indicates the azimuth or orientation

**Figure ISM8** High resolution, 3D landscapes of Neolithic Stenness, Orkney**.** Created with the software *Horizon* by A.G.K. Smith, © A.G.K. Smith. Created with Terrain 50. Contains Ordnance Survey data Crown copyright and database right (2012). http://www.ordnancesurvey.co.uk/docs/licenses/os-opendata-licence.pdf

**Figure ISM9** High resolution, 3D landscapes of Neolithic Callanish, Lewis. Created with the software *Horizon* by A.G.K. Smith, © A.G.K. Smith. Created with Terrain 50. Contains Ordnance Survey data Crown copyright and database right (2012). http://www.ordnancesurvey.co.uk/docs/licenses/os-opendata-licence.pdf

# Supplementary Material

### 8.1. Understanding the Sun and Moon's Declination

The Sun's declination varies throughout the year, reaching a maximum northerly value (23.5° above the celestial equator) at the summer solstice (June 20–21) and a maximum southerly value (23.5° below the celestial equator) at the winter solstice (December 20–21). The declination is zero when it crosses the celestial equator at the equinoxes (March 20 and September 22). An observer on the ground will see the Sunrise *more or less* due east and set *more or less* due west at the equinoxes, rise and set at its most northerly points at the summer solstice and at its most southerly points at the winter solstice. Further, an observed physical mid-point between the solstices on the horizon is unlikely to equal the actual time of the Equinox when the Sun crosses the celestial equator moving northwards or southwards (celestial equator: an extension of the Earth's equator drawn out into the celestial sphere).

The Moon's declination also varies throughout its monthly cycle, in much the same way as the Sun's over the course of a year. Since the Moon's path is basically similar to that of the Sun, its maximum declination from the celestial equator is about 23.5° on average. However, the points at which it reaches its maximum declination and turns around are known as lunistices, by analogy with the solstices of the Sun. Since the Moon's orbit is inclined at about 5° to the ecliptic, the declinations of the lunistices vary up to 5° above and below the average value, i.e. from about 18.5° to 28.5°. This variation occurs on a cycle with a period of about 18.6 years. The time at which the minimum value occurs is known as the Minor Lunar Standstill or the minor standstill of the Moon and that of the maximum value as the Major Lunar Standstill or the major standstill of the Moon. Visually, over the course of a month, the Moon appears to rise in a progressively northerly direction until it reaches an extreme point, then turns around and rises in a progressively southerly direction until it reaches its southerly extreme, then turns back to the north again. The range between these extreme points is greater at the major lunar standstill than at the minor lunar standstill. The Minor Lunar Standstill can occur twice a month and is when the Moon appears to be travelling within the bands of the Sun's paths *between* the Summer and Winter Solstices (see the green lines on the various 3D landscapes), whereas at the Major Lunar Standstill the Moon is seen to travel out-with these bands (see the red lines on the various 3D landscapes),

Finally, a minus sign in front of a declination figure means the phenomena is in the south, such as −28.5°; no sign or a '+' sign means it occurs in the north.

### 8.2. Understanding the Physical Dynamics of the Celestial Bodies Observable at the Monuments

In agreement with Martlew and Ruggles (1993), it is the full Moon that was the most likely candidate for the lunar alignments given the celestial dynamics and the resulting celestial show that occurs at these times. There is no simple relationship between the solar and lunar cycles, to the extent that at a solstice the phase of the Moon varies from year to year, although a full Moon must always occur within 2 weeks of a solstice and, since the Sun is moving slowly around the time of the solstice, the celestial dynamics of the Sun and the Moon remain largely unchanged throughout this period. If the Sun is at or near the summer solstice, i.e. in the north, then the full Moon, which must be directly opposite the Sun, must be in the south in the vicinity of the winter solstice point. It is in this direction that most distant horizons occur and monument orientations face. The converse is also true, when the Sun is at or near the winter solstice point in the south, the full Moon must be near the summer solstice point in the north: that is, where the majority of closest horizons occur and the direction of three orientations.

It is also important that as the time of the solstice draws near the difference between the Sun's rising positions along the horizon daily decreases, until it appears to stop, although reference points on the horizon, such as peaks, notches and trees, will help an observer to track the Sun's rising. The same is true of the Sun's setting points. Eventually, the Sun stops and then changes direction, returning almost imperceptibly at first and then it can be seen rising or setting at greater intervals along the horizon. On Mull and in Argyll, there are orientations, sometimes one of a pair, that actually indicate the solstices (Higginbottom 2003; GH 4 *in preparation*; Ruggles 1984).

The monuments on Mull and in Argyll clearly highlight the cosmic order of opposites: at the extremes, the Sun is on view for the longest time possible in summer during the day and the full Moon is on display for the shortest at night, while in winter the opposite is true.

### 8.3. The 3D Horizon program settings

The dates have been set to 1500 BC for the Bronze Age standing stones sites and 3000 BC for the two Neolithic Great Circles of Callanish and Stenness. The observation height is taken to be 1.5 metres. The z-co-ordinate for elevation for ISMFig. 1 & 3-7 has been multiplied by 1.5. These figures were used to make the composite figure in the text of the article (Fig. 3). This magnification was necessary to enable clear viewing of the horizons in the highly reduced composite figure. Note that the z-co-ordinate of Dunamuck was multiplied by 1.5 for the composite figure only (Fig. 3b), but not for its 3D landscape used for ISMFig2.

The amount of 'haze' or 'fog' in the atmosphere can be simulated through a 'Atmospheric Clarity' variable in the *Horizon* program by Smith. This is useful "because some haze provides an illusion of depth in the rendered image, as well as some indication of what a real observer might actually be seeing. It is also used to determine the extinction coefficient for stars. In reality, the brightness of a landscape feature falls off exponentially with distance, and gradually merges with the blueness of the sky. The optical depth of the atmosphere, *i.e.* the distance in which the brightness falls to about 37% of its original value, is specified through the Distance (km) option (A.G.K. Smith, 2013, *Horizon* User Guide and Implementation Notes. Documentation Version 0.12 December 3, 2013: 13). Whilst a typical value is about 10 to 20 kilometres, and whilst it can vary from a few kilometres up to over 100 kilometres, we have chosen 75 km. We also ran the *Horizon* program at full-resolution as well as in the creation of the output files. The advantage of doing these things for the publication was to create horizon profiles that were as sharp and clear as possible so they would not lose their clarity when they were reduced. Using 75 rather than 100 km helped to make the pictures a little less harsh. However, there is some lack of smoothing as a result which is noticeable in some of these ISM figures.

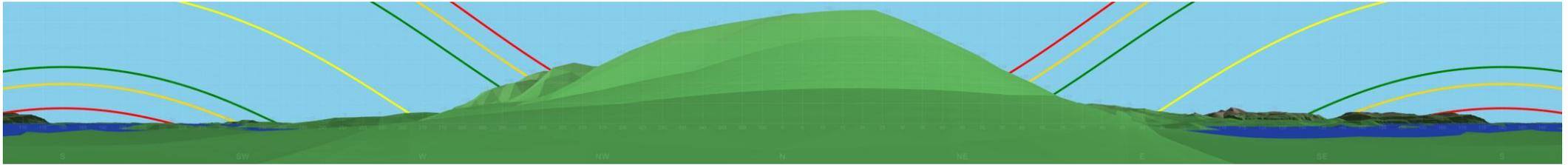

**Figure ISM1** High resolution, 3D landscapes of Cragaig, Mull. Created with the software *Horizon* by A.G.K. Smith, © A.G.K. Smith. Created with Terrain 50. Contains Ordnance Survey data Crown copyright and database right (2012). http://www.ordnancesurvey.co.uk/docs/licenses/os-opendata-licence.pdf. Data and licencing information and copyright the same for all *in-line* landscapes below. **For colour reproduction on the Web.**

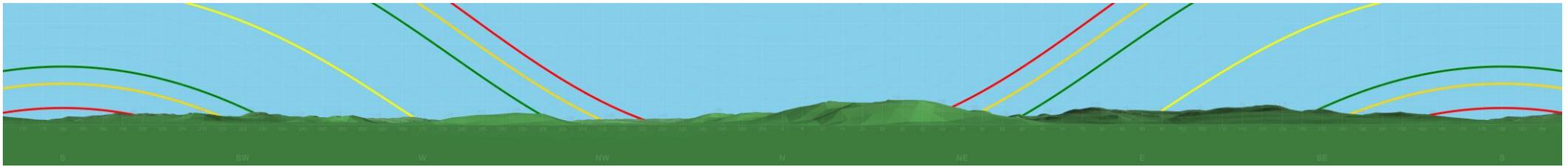

**Figure ISM2** High resolution, 3D landscapes of Dunamuck, Argyll. Created with the software *Horizon* by A.G.K. Smith, © A.G.K. Smith. **For colour reproduction on the Web.**

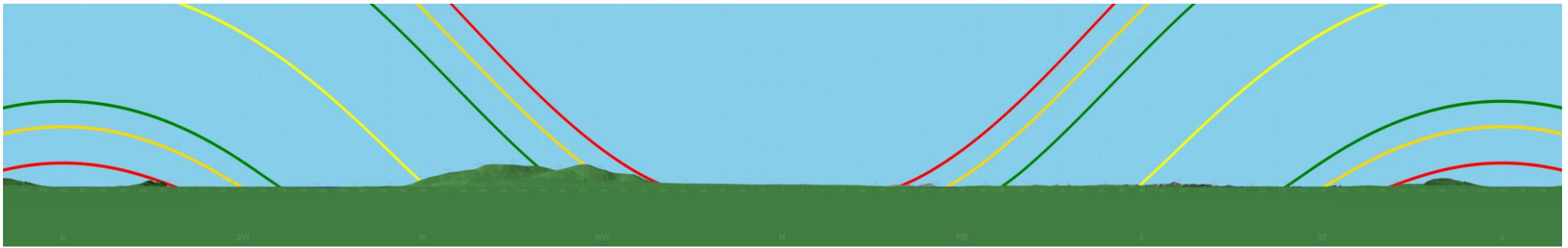

**Figure ISM3** High resolution, 3D landscapes of Hough, Tiree. Ratio 1:1.5 width:height, due to low hoizons. Created with the software *Horizon* by A.G.K. Smith, © A.G.K. Smith. **For colour reproduction on the Web.**

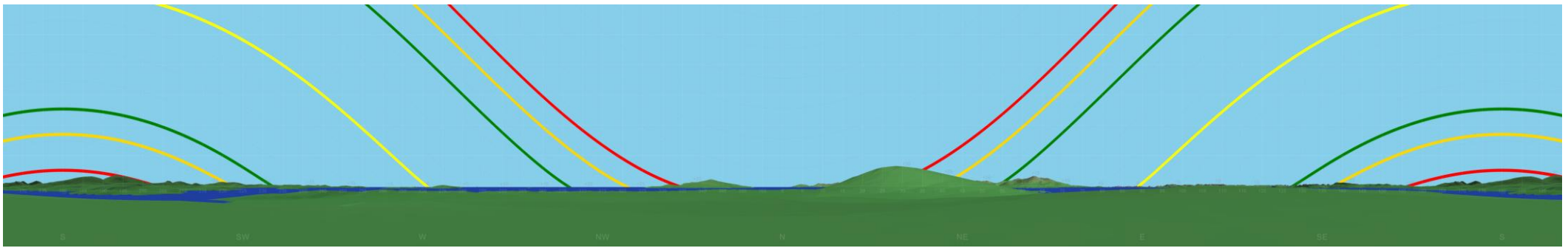

**Figure ISM4** High resolution, 3D landscapes of Blashavel, Uist. Created with the software *Horizon* by A.G.K. Smith, © A.G.K. Smith. **For colour reproduction on the Web.**

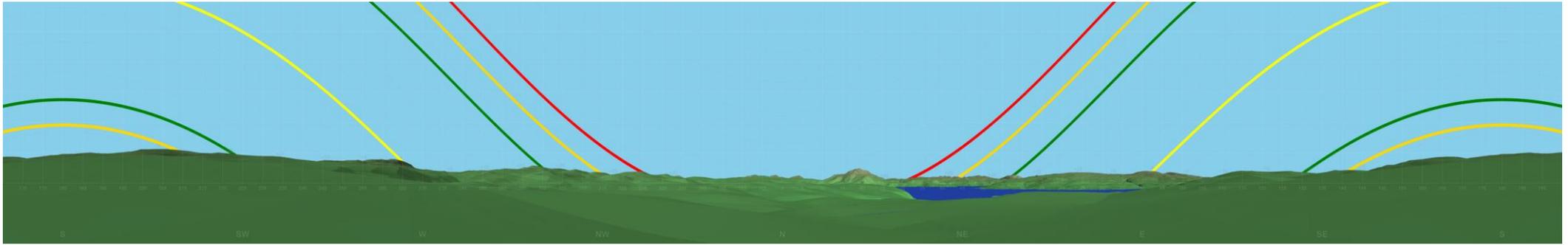

**Figure ISM5** High resolution, 3D landscapes of Balliscate, Mull, a reverse site**.** Created with the software *Horizon* by A.G.K. Smith, © A.G.K. Smith. **For colour reproduction on the Web.**

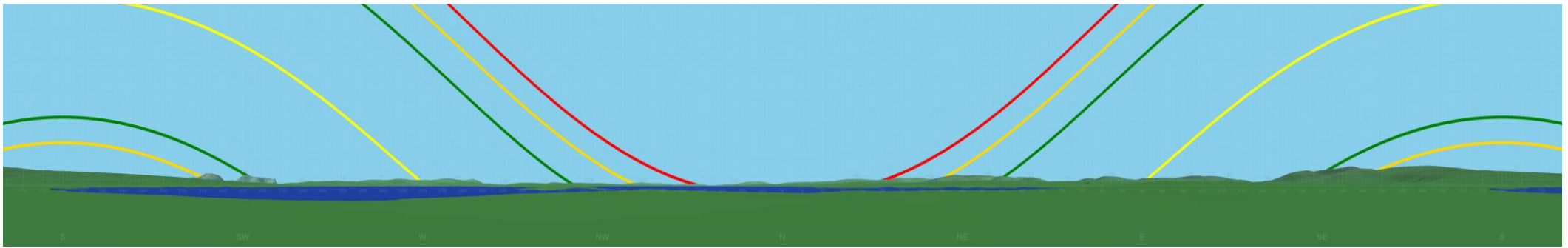

**Figure ISM8** High resolution, 3D landscapes of Neolithic Stenness, Orkney, a reverse site**.** Created with the software *Horizon* by A.G.K. Smith, © A.G.K. Smith. Created with Terrain 50. Contains Ordnance Survey data Crown copyright and database right (2012). http://www.ordnancesurvey.co.uk/docs/licenses/os-opendata-licence.pdf  **For colour reproduction on the Web.**

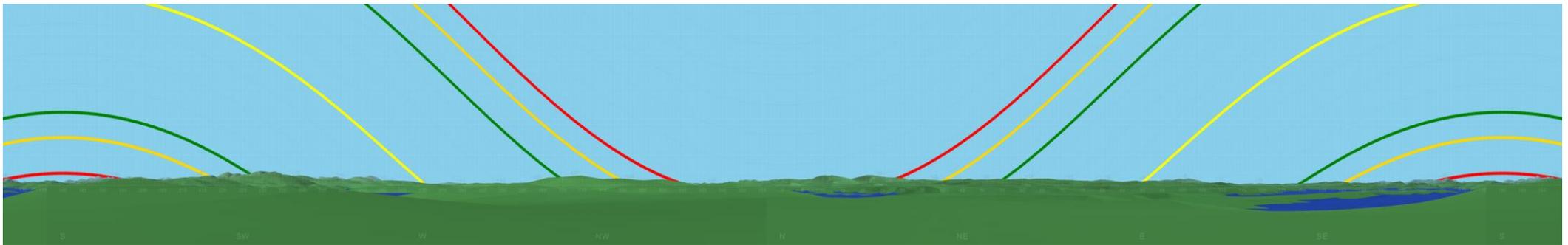

**Figure ISM9** High resolution, 3D landscapes of Neolithic Callanish, Lewis. Created with the software *Horizon* by A.G.K. Smith, © A.G.K. Smith. Created with Terrain 50. Contains Ordnance Survey data Crown copyright and database right (2012). http://www.ordnancesurvey.co.uk/docs/licenses/os-opendata-licence.pdf  **For colour reproduction on the Web.**

29